\documentclass[aps, prx, twocolumn, showpacs, superscriptaddress]{revtex4-2}
\usepackage{graphicx}
\usepackage{amsmath}
\usepackage{mathtools}
\usepackage{dsfont}
\usepackage{amssymb}
\usepackage{physics}
\usepackage{hyperref}
\usepackage{ulem}
\usepackage{xcolor}
\usepackage{xr} 

\hypersetup{colorlinks=true,
	    final=true,
	    linkcolor=blue,
	    citecolor=blue,
	    filecolor=blue,
	    urlcolor=blue,
}

\newcommand{\omax}{\omega_{\max}}
\newcommand{\wh}[1]{\widehat{#1}}

\DeclareMathOperator*{\argmin}{arg\,min}

\begin{document}
\title{Stabilizing the calculation of the self-energy in dynamical mean-field theory using constrained residual minimization}
\author{Harrison LaBollita}
\affiliation{Department of Physics, Arizona State University, Tempe, AZ 85287, USA}
\affiliation{Center for Computational Quantum Physics, Flatiron Institute, 162 5th Avenue, New York, NY 10010, USA}
\author{Jason Kaye}
\affiliation{Center for Computational Quantum Physics, Flatiron Institute, 162 5th Avenue, New York, NY 10010, USA}
\affiliation{Center for Computational Mathematics, Flatiron Institute, 162 5th Avenue, New York, NY 10010, USA}
\author{Alexander Hampel}
\affiliation{Center for Computational Quantum Physics, Flatiron Institute, 162 5th Avenue, New York, NY 10010, USA}

\begin{abstract}
We propose a simple and efficient method to calculate the electronic self-energy in dynamical mean-field theory (DMFT), addressing a numerical instability often encountered when solving the Dyson equation. Our approach formulates the Dyson equation as a constrained optimization problem with a simple quadratic objective. The constraints on the self-energy are obtained via direct measurement of the leading order terms of its asymptotic expansion within a continuous time quantum Monte Carlo framework, and the use of the compact discrete Lehmann representation of the self-energy yields an optimization problem in a modest number of unknowns. We benchmark our method for the non-interacting Bethe lattice, as well as DMFT calculations for both model systems and {\it ab-initio} applications. 

\end{abstract}

\maketitle

\section{\label{sec:introduction}introduction}
Quantum impurity models, which generically describe a quantum impurity interacting with a non-interacting fermionic bath \cite{Anderson1961}, play an important role in the study of strongly correlated materials \cite{Gull2011continuous}. In particular, they serve as auxiliary models in quantum embedding techniques such as dynamical mean-field theory (DMFT) used to map an interacting lattice problem to a quantum impurity problem subject to a self-consistency condition \cite{georges1996dynamical}. 
A common choice of quantum impurity solver is continuous-time quantum Monte Carlo (CT-QMC) using the hybridization expansion algorithm (CT-HYB) \cite{werner2006hybridization,Gull2011continuous,Nolting1972Metho}, in which observables of the quantum impurity, such as the impurity Green's function, the many-body impurity density matrix, and higher-order correlation functions, are directly measured via sampling.

In DMFT and its {\it ab-initio} extension density functional theory (DFT) + DMFT, the impurity Green's function $G$ and its corresponding electronic self-energy $\Sigma$, matrices with the dimension of the impurity space, are the primary quantities of interest. Within the DMFT formalism, these quantities are related via the Dyson equation, formulated in the Matsubara frequency domain as
\begin{equation} \label{eq:dyson1}
    G(i\nu_n) = G_0(i\nu_n) + G_0(i\nu_n) \Sigma(i\nu_n) G(i\nu_n),
\end{equation}
or equivalently
\begin{equation}
\label{eqn:dyson}
    \Sigma(i\nu_n) = G_{0}^{-1}(i\nu_{n})- G^{-1}(i\nu_{n}).
\end{equation}
Here, $\nu_{n} \equiv (2n+1) \pi / \beta$ is a fermionic Matsubara frequency at inverse temperature $\beta$, and $G_{0}$ is a bare Green's function, which contains the dynamical Weiss field describing the impurity bath coupling and the non-interacting part of the impurity Hamiltonian. It is well-known that computing the self-energy using the direct formula \eqref{eqn:dyson} leads to an amplification of noise, typically arising from Monte Carlo-based impurity solvers, which grows with increasing $\nu_n$ \cite{Gull2011continuous}. The mechanism of this phenomenon will be discussed in this article. Since high-frequency information contributes significantly to quantities, such as the Weiss field, required to satisfy the DMFT self-consistency condition, excessive noise ultimately prevents the convergence of the DMFT loop and charge self-consistency within DFT+DMFT. Several methods are used to address this issue, the most common of which we briefly summarize.

An approach referred to as ``tail-fitting'' attempts to address the problem directly \cite{Haule2010, Priyanka2016cthyb}. The coefficients of a high-frequency asymptotic expansion $\Sigma(i\nu_{n}) \approx \sum_{j=0}^N \Sigma_j/(i\nu_{n})^j$ of the self-energy are fit from data obtained using \eqref{eqn:dyson} in some fitting window ($\nu_{n}^{\mathrm{min}}$, $\nu_{n}^{\mathrm{max}}$). The window must be chosen to cover sufficiently high frequencies so that the asymptotic expansion is valid, but not so high that the data for the fit is overwhelmed by noise. This asymptotic expansion is then used in place of \eqref{eqn:dyson} at high frequencies. Determining the location and width of the fitting window, the number of terms to include in the asymptotic expansion, and the point at which to ``glue'' the asymptotic expansion to the representation \eqref{eqn:dyson} requires substantial tuning on a case-by-case basis. This often makes it difficult to obtain well-converged DMFT results, and represents a significant obstacle to the construction of reliable, black-box DMFT codes.

Another approach is to expand the imaginary time Green's function in a basis of orthogonal polynomials, such as Legendre polynomials \cite{Boehnke2011Legendre}, constraining the expansion coefficients using the known high-frequency asymptotics of the Green's function (in particular, $G(i\nu_{n}) \sim 1/(i\nu_n)$ as $n \to \infty$). The expansion coefficients can be directly measured, at some computational expense, within the CT-HYB algorithm. Enforcing the correct leading-order asymptotic behavior of the Green's function reduces, but does not eliminate, the amplification of high-frequency noise in the self-energy. The performance of this approach is also sensitive to the size of the basis, which is challenging to select in a manner that fits the data accurately without overfitting to noise.

A third approach, based on the equations of motion, is the method of improved estimators (IE) \cite{Hafermann2012improved, Hafermann2014, Gunacker2016worm}, and more recently symmetric improved estimators (sIE) \cite{Kaufmann2019,Kugler2022}. These involve the derivation of higher-order correlation functions related to the self-energy (and vertex) corrections, which can be directly measured within CT-HYB \cite{Hafermann2012improved}. While the sIE approach is effective in stabilizing the self-energy calculation, it requires some implementation effort and a significant increase in computational cost, since unlike the impurity Green's function in CT-HYB, the sIEs cannot be measured efficiently by partition sum sampling~\cite{Kaufmann2019}.

Our approach is based on the observation that the problem of high frequency noise amplification in the self-energy is simply the result of a numerical instability in \eqref{eqn:dyson}: since both $G_{0}^{-1}(i \nu_n)$ and  $G^{-1}(i \nu_n)$ grow with $\nu_n$, but $\Sigma(i \nu_n)$ does not, the subtraction causes a loss of accuracy due to catastrophic cancellation. To address this, we therefore avoid \eqref{eqn:dyson} and instead obtain $\Sigma$ by using \eqref{eq:dyson1} to set up a constrained minimization problem: minimize the residual $\norm{G - G_0\Sigma G - G_0}$, in a suitable norm $\norm{\cdot}$, subject to a constraint on the high-frequency behavior of $\Sigma$. These high-frequency asymptotics can be written in terms of expectation values of equal-time operators, which can be straightforwardly obtained using any CT-QMC-based solver. Discretizing all quantities using the compact discrete Lehmann representation (DLR) \cite{Kaye2022DLR} leads to a quadratic minimization problem with a scalar constraint in a small number of degrees of freedom, even at very low temperatures and large spectral widths. The resulting \textit{constrained residual minimization} (CRM) scheme is simple to implement using existing codes, with a negligible computational cost. In our numerical experiments using the CT-HYB solver, we find robust convergence of the self-energy with no tuning of parameters, leading to accuracy at the level of the Monte Carlo noise with the correct high-frequency behavior.

We note the related method of Ref. \cite{han21}, a constrained optimization-based fitting procedure combining directly computed self-energy data up to a specified frequency cutoff (as in tail-fitting methods) with independently computed asymptotic coefficients of $\Sigma(i \nu_n)$. We show that in addition to the computed asymptotic coefficients, using the residual of the Dyson equation as an objective function and the DLR parameterization eliminates the need for a manually-chosen frequency cutoff.

This paper is organized as follows. After introducing the required numerical tools in Sec. \ref{sec:prelim} and describing our method in Secs. \ref{sec:instability} and \ref{sec:crm}, we validate our method for the exactly solvable non-interacting single-band Bethe lattice in Sec. \ref{sec:nibethe}. In Sec. \ref{sec:results}, we then demonstrate the full procedure for two representative examples: the interacting Bethe lattice, and an {\it ab-initio} DMFT calculation for Sr$_{2}$RuO$_{4}$.

\section{\label{sec:prelim}Preliminaries}
Our method relies on two primary tools: the high-frequency expansion of the self-energy, and a compact representation of imaginary time and imaginary frequency quantities called the DLR.

\subsection{\label{sec:hfreq}High-frequency expansion of the self-energy}

A generic quantum impurity model Hamiltonian is given by
\begin{equation}
\label{eq:H}
\mathcal{H} = \mathcal{H}_{0} + \mathcal{H}_{\mathrm{int}},
\end{equation}
with 
\begin{multline*}
\mathcal{H}_{0} = \sum_{k\alpha} \varepsilon_{k\alpha}c^{\dagger}_{k\alpha}c_{k\alpha}\\ + \sum_{k\alpha b} \big (V_{k}^{\alpha b}c^{\dagger}_{k\alpha}d_{b} + \mathrm{h.c.} \big ) + \sum_{ab}E_{ab}d^{\dagger}_{a}d_{b}
\end{multline*}
and
\begin{equation*}
\mathcal{H}_{\mathrm{int}} = \frac{1}{2}\sum_{abcd}U_{abcd}d^{\dagger}_{a}d^{\dagger}_{b}d_{d}d_{c}.
\end{equation*}
Here, $c^{\dagger}_{\alpha}$ and $d^{\dagger}_{a}$  are the fermion creation operators in the bath and impurity, respectively. $E_{ab}$ captures the bare energy structure for the quantum impurity, $U_{abcd}$ parameterizes the electron-electron interactions, $V_{k}^{\alpha b}$ describes the hybridization between the impurity and the bath, and $\varepsilon_{k \alpha}$ is the energy-momentum dispersion of the bath.

The electronic self-energy of the quantum impurity problem can be expanded in the $\nu_n \to \infty$ limit as
\begin{equation} \label{eq:sigasymp}
\Sigma(i\nu_{n}) = \Sigma_{0} + \frac{\Sigma_{1}}{i\nu_{n}} + \mathcal{O}\big(\nu_{n}^{-2}\big),
\end{equation}
where the matrices $\Sigma_0$ and $\Sigma_1$ can be obtained from the high-frequency expansions of the full and non-interacting Green's functions. 
They are formally defined in terms of expectation values of commutators and anti-commutators,
\begin{equation} \label{eq:moments}
\begin{aligned}
    \Sigma_{0}^{ab} &= - \langle \{ [\mathcal{H}_{\mathrm{int}}, d_{a}], d_{b}^{\dagger}\} \rangle\\
    \Sigma_{1}^{ab} &= \langle \{ [\mathcal{H}_{\mathrm{int}}, [\mathcal{H}_{\mathrm{int}}, d_{a}]], d_{b}^{\dagger}\rangle - (\Sigma_{0}^{ab})^{2},
\end{aligned}
\end{equation}
which can be computed to high accuracy within CT-HYB from the many-body impurity density matrix at no additional computational cost  \cite{Wang2011high, Gull2011continuous, Haule2007CTQMC}. Formally, the full Hamiltonian \eqref{eq:H} should be used in the commutators, but all non-trivial terms from its non-interacting part cancel in $\Sigma_0$ and $\Sigma_1$. 

The high-level abstractions used in the TRIQS software library~\cite{Parcollet2015triqs} allow for the direct implementation of the commutators in \eqref{eq:moments}. Such an implementation is general, and in particular agnostic to the exact form of $\mathcal{H}_{\mathrm{int}}$. Further details on the derivation and implementation of \eqref{eq:moments} are given in Appendix~\ref{sec:app_high_freq}.

\subsection{\label{sec:dlr}Discrete Lehmann representation}
We give a brief overview of the DLR, and refer the reader to Refs.~\onlinecite{Kaye2022DLR,kaye22_libdlr} for further details. The DLR of an imaginary time Green's function is given by an expansion of the form
\begin{equation} \label{eq:dlrexpansion}
    G(\tau) \approx \sum_{l=1}^r K(\tau, \omega_l) \, \widehat{g}_l,
\end{equation}
where $K(\tau,\omega) = -\frac{e^{-\tau \omega}}{1+e^{-\beta \omega}}$ is the analytic continuation kernel appearing in the spectral Lehmann representation
\begin{equation} \label{eq:lehmann}
    G(\tau) = \int_{-\infty}^\infty d\omega \, K(\tau, \omega) \rho(\omega)
\end{equation}
relating $G$ to its spectral density $\rho$. 
The DLR frequencies $\omega_l$ are obtained by applying a rank-revealing pivoted Gram-Schmidt algorithm to the analytic continuation kernel, and are therefore independent of the specific Green's function $G$. They depend only on the desired accuracy $\epsilon$ of the DLR expansion \eqref{eq:dlrexpansion}, and a dimensionless cutoff parameter $\Lambda = \beta \omax$, where $\omax$ is such that $\rho(\omega) = 0$ outside $[-\omax, \omax]$. Both $\epsilon$ and $\Lambda$ are user-specified parameters. To determine $\Lambda$, typically $\beta$ is given, and $\omax$ can be estimated, or results can be converged with respect to $\Lambda$. For any imaginary time Green's function $G$ obeying the cutoff $\Lambda$, there are coefficients $\widehat{g}_l$ such that \eqref{eq:dlrexpansion} holds to accuracy $\epsilon$; in other words, the functions $K(\tau,\omega_l)$ span the space of all such Green's functions. The existence of the DLR expansion is derived from a low-rank approximation of the analytic continuation operator \eqref{eq:lehmann}, and crucially, it is observed that $r = \mathcal{O}(\log(\Lambda) \log(\epsilon^{-1}))$, yielding an exceptionally compact representation.

The DLR coefficients $\widehat{g}_l$ can be determined by fitting the expansion \eqref{eq:dlrexpansion} to data, or by interpolation at a collection of $r$ DLR nodes $\tau_k$ also obtained using the pivoted Gram-Schmidt algorithm. The latter requires solving an $r \times r$ linear system obtained by evaluating the expression \eqref{eq:dlrexpansion} at $\tau = \tau_k$.
Fourier transform of \eqref{eq:dlrexpansion} yields a DLR expansion
\begin{equation} \label{eq:dlrexpansionmf}
    G(i \nu_n) \approx \sum_{l=1}^r K(i \nu_n, \omega_l) \, \widehat{g}_l
\end{equation}
in the Matsubara frequency domain, with $K(i \nu_n, \omega) = (i\nu_n - \omega)^{-1}$ for fermionic Green's functions. As in imaginary time, the coefficients $\widehat{g}_l$ can be determined by interpolation at a collection of $r$ DLR nodes $i \nu_{n_k}$. Once the DLR coefficients $\widehat{g}_l$ have been obtained, the DLR expansion can be evaluated both in the imaginary time and frequency domains, so the Fourier transform is in effect performed analytically.

We note that the DLR is a close cousin of the intermediate representation (IR) \cite{shinaoka17,chikano18}, which uses a singular value decomposition of the analytic continuation kernel to obtain an orthogonal expansion in terms of numerically-represented basis functions, rather than the DLR's non-orthogonal expansion in terms of analytically-known basis functions. The number of terms in the DLR and IR expansions at given $\Lambda$ and $\epsilon$ is similar, with the IR containing slightly fewer. An IR expansion can also be obtained using fitting or a similar interpolation procedure, with interpolation nodes determined either in the manner described above \cite{Kaye2022DLR}, or using the sparse sampling method \cite{li20,shinaoka22}. Both the DLR and IR have been effectively used to represent imaginary time and frequency quantities other than Green's functions, such as self-energies and hybridization functions \cite{li20,kaye23_eqdyson,yeh22,cai22,shinaoka22,Sheng2023,kaye23_diagrams}.

\section{\label{sec:algo}Constrained residual minimization algorithm}

\subsection{\label{sec:instability}Instability in the direct solution of the Dyson equation} 

The formula \eqref{eqn:dyson} for $\Sigma$ is numerically unstable, causing an amplification of noise in $G_0$ and $G$ at high frequencies. To demonstrate this, suppose we perturb $G(i \nu_n)$ by a noise function $\eta(i \nu_n)$, and let $\widetilde{\Sigma}$ be the approximation of $\Sigma$ obtained from \eqref{eqn:dyson}. We find
\begin{align*}
    \widetilde{\Sigma}(i \nu_n) &= G_0(i \nu_n)^{-1} - (G(i \nu_n) + \eta(i \nu_n))^{-1} \\
    &= G_0(i \nu_n)^{-1} - G(i \nu_n)^{-1} (I +  \eta(i \nu_n) G(i \nu_n)^{-1})^{-1} \\
    &= G_0(i \nu_n)^{-1} - G(i \nu_n)^{-1} \sum_{k=0}^\infty (-1)^{k}(\eta(i \nu_n) G(i \nu_n)^{-1})^k \\
    &= \Sigma(i \nu_n) - G(i \nu_n)^{-1} \sum_{k=1}^\infty (-1)^{k}(\eta(i \nu_n) G(i \nu_n)^{-1})^k \label{eqn:sigma_tilde}
\end{align*}
as long as $\eta$ is sufficiently small. We focus on the case of scalar-valued Green's function and self-energy for simplicity.
Since $G(i \nu_n) \sim (i \nu_n)^{-1}$ as $n \to \infty$, we have $G(i \nu_n)^{-1} \sim i \nu_n$, yielding an expansion of the error of the form
\[\eta \nu_n^2 + \eta^2 \nu_n^3 + \cdots.\]
Here we have abused notation and ignored the frequency dependence of $\eta$. This error growth can be viewed as a consequence of catastrophic cancellation: the growing quantities $G_0^{-1}$ and $G^{-1}$ are subtracted to obtain a bounded quantity $\Sigma$, leading to an amplification of noise. We note that while this explanation assumes roughly uniform noise in the Matsubara frequency domain, we do not claim this is necessarily the correct error model for any given Monte Carlo-based impurity solver. However, it is a simple model of the mechanism of high frequency noise amplification, which reproduces the rate of noise growth often observed in practice.

\subsection{\label{sec:crm}Constrained residual minimization}
To avoid the numerically unstable formula \eqref{eqn:dyson}, we instead begin with \eqref{eq:dyson1}, defining its residual as 
\begin{equation}
    \label{eq:res1}
    R = G - G_{0}\Sigma G- G_{0}. 
\end{equation}

Here, all quantities are considered to be numerical approximations, so that $R \neq 0$ as it would be if \eqref{eq:dyson1} were satisfied exactly. Given the leading coefficients $\Sigma_0$ and $\Sigma_1$ of the asymptotic expansion \eqref{eq:sigasymp}, which can be computed directly as described above, $\Sigma$ can be obtained from $G$ and $G_0$ by minimizing $\norm{R}$ in a suitable norm subject to the constraint that \eqref{eq:sigasymp} holds. We refer to this approach as constrained residual minimization (CRM).

We discretize \eqref{eq:res1} by evaluating it at the DLR Matsubara frequency nodes $i \nu_{n_k}$. Thus, we define $r_k \equiv R(i \nu_{n_k})$, $g_{k}\equiv G(i\nu_{n_{k}})$, $g_{0k}\equiv G_{0}(i\nu_{n_{k}})$, and $\sigma_{k}\equiv \Sigma(i\nu_{n_{k}})$. As described above, the DLR expansions of $R$, $G$, and $G_0$ can be recovered from the values $g_k$ and $g_{0k}$. For $\Sigma$, we must first subtract its known constant part $\Sigma_{0}$ for the DLR expansion to be valid. Thus, defining $\overline{\Sigma} \equiv \Sigma - \Sigma_0$, we can recover the DLR expansion of $\overline{\Sigma}$ from the values $\sigma_k - \Sigma_0$, which gives a method of recovering $\Sigma$, and also implements the constraint on the constant term given in \eqref{eq:sigasymp}. Our approach is similar to the method of Legendre polynomials described above; however, we apply the constraint directly to the self-energy. The discretization of \eqref{eq:res1} is therefore given by
\begin{equation} \label{eq:resdisc}
    r_{k} \equiv g_{k}-g_{0_{k}}\sigma_{k}g_{k} - g_{0_{k}}.
\end{equation}

\begin{figure*}
    \centering
    \includegraphics[width=2\columnwidth]{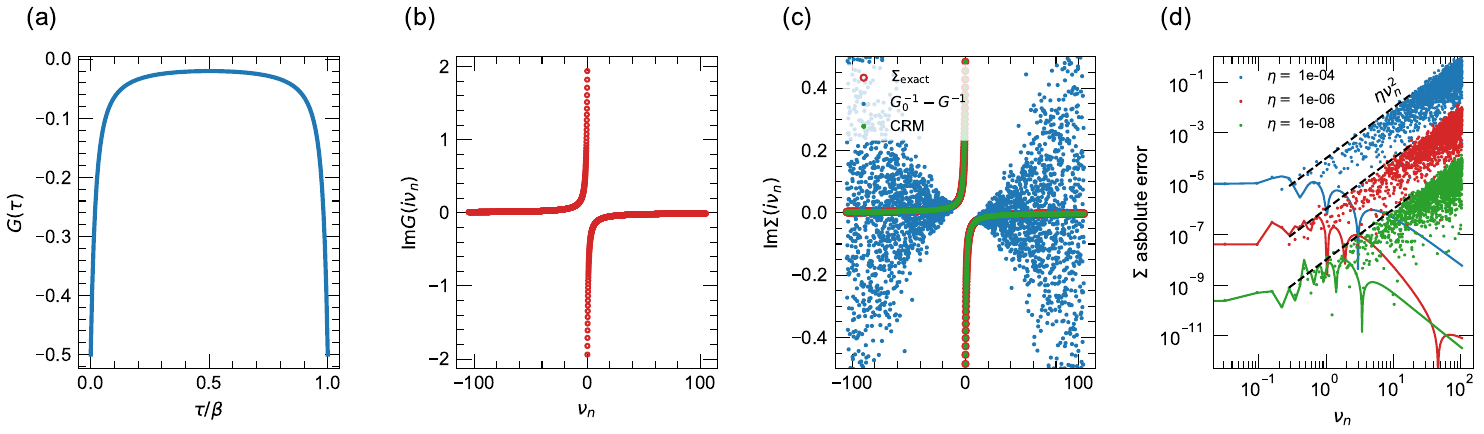}
    \caption{(a) $G(\tau)$ for the non-interacting Bethe lattice with $\beta D= 100$. (b) $\Im G(i\nu_{n})$; note $\text{Re} \, G(i \nu_n) = 0$ by symmetry. (c) $\Im \Sigma(i\nu_{n})$ computed via the direct formula \eqref{eqn:dyson} and the CRM method, using data obtained in imaginary time with noise level $\eta = 10^{-4}$, along with the exact solution $\Sigma = G/4$.  (d) Pointwise error for the direct method (dots) and the CRM method (solid line), for several choices of $\eta$. Black dashed line denotes $\eta\nu_{n}^{2}$.}
    \label{fig:exactcase}
\end{figure*}

We measure the residual \eqref{eq:res1} in the $L^2(\tau)$ norm, given by
\begin{equation} \label{eq:rres}
    \norm{R}_{L^2(\tau)}^2 \equiv \frac{1}{\beta} \int_0^\beta d\tau \, \norm{R(\tau)}_F^2,
\end{equation}
where $\norm{\cdot}_F$ is the Frobenius norm.
We show in Appendix~\ref{app:l2norm} that \eqref{eq:rres} can be computed from a simple quadratic form evaluated at the DLR coefficients $\widehat{r}_l$ of $R$, which can be obtained from the values $r_k$ by interpolation.

To implement the constraint on the first decaying term in \eqref{eq:sigasymp}, we equate the DLR expansion of $\overline{\Sigma}$ and the desired asymptotic expansion:
\[
\begin{multlined}
\sum_{l=1}^r K(i \nu_n, \omega_l) \widehat{\sigma}_l = \overline{\Sigma}(i \nu_n) \\ = \Sigma(i \nu_n) - \Sigma_0 = \frac{\Sigma_1}{i \nu_n} + \mathcal{O}(\nu_n^{-2}).
\end{multlined}
\]
Multiplying both sides by $i \nu_n$ and taking the limit $n \to \infty$ yields the desired constraint 
\[\sum_{l=1}^r \widehat{\sigma}_l = \Sigma_1,\]
since $K(i \nu_n,\omega) = (i \nu_n - \omega)^{-1}$.

The discretized CRM problem is therefore given as follows:
\begin{equation} \label{eq:minimdisc}
 \argmin_{\sigma_k} \norm{R}_{L^2(\tau)} \quad \text{subject to} \quad \sum_{l=1}^r \widehat{\sigma}_l = \Sigma_1.
\end{equation}
Once it has been solved, $\Sigma$ is given by the DLR expansion
\[\Sigma(i \nu_n) = \Sigma_0 + \sum_{l=1}^r K(i \nu_n, \omega_l) \widehat{\sigma}_l.\]
The objective function $\norm{R}_{L^2(\tau)}$ can be computed from $\sigma_k$ using \eqref{eq:resdisc} and the procedure described in Appendix \ref{app:l2norm}, since $g_k$ and $g_{0k}$ are given. The DLR coefficients $\widehat{\sigma}_l$ are recovered from the $\sigma_k$ by solving the $r \times r$ linear system
\[\sum_{l=1}^r K(i \nu_{n_k}, \omega_l) \widehat{\sigma}_l = \sigma_k.\]

In our numerical examples, we use the SciPy \cite{2020SciPy-NMeth} implementation of the trust region optimization method \cite{Conn2000trust} to solve \eqref{eq:minimdisc}.
We note that the CRM problem is simply a quadratic minimization problem with a scalar constraint, and its solution is therefore given semi-analytically in terms of simple matrix operations. However, we find that the trust region optimization tends to perform slightly better, due to the appearance of ill-conditioned operations in the semi-analytic formula.

We note that since the CRM scheme seeks a balance between correctly matching the low-frequency data via the Dyson equation, and high-frequency asymptotics of relatively low-order, one might obtain a larger error for intermediate frequencies, particularly at low temperatures, in which neither of these limiting representations is sufficiently accurate. In our numerical examples, which include realistic DMFT calculations, we have not observed a significant contribution of this effect to the error (including low temperature calculations). A topic of our future work is the possibility of directly measuring higher-order asymptotics, which would push the region of validity of the asymptotic expansion to lower frequencies. In particular, we note that the next term of the asymptotic expansion can be obtained given an approximation of the bath hybridization by a finite number of levels, for which there exist a variety of algorithms \cite{georges1996dynamical,mejutazaera20,shinaoka21,huang23}.

\subsection{\label{sec:nibethe}Benchmark: non-interacting Bethe lattice}
We first test our algorithm on the non-interacting Bethe lattice, given by \eqref{eq:dyson1} with $\Sigma = G/4$ and $G_0(i \nu_n) = (i \nu_n)^{-1}$. This model can be solved analytically in real frequency \cite{georges1996dynamical}, yielding the semi-circular spectral function $\rho(\omega) = \frac{2}{\pi}\sqrt{1-\omega^2}$ on $[-1,1]$. $G(\tau)$ and $G(i \nu_n)$ can then be computed from the Lehmann representation \eqref{eq:lehmann} and its Fourier transform (obtained by replacing $K(\tau,\omega)$ with $K(i \nu_n, \omega)$) by numerical integration. These are shown in Figs. \ref{fig:exactcase}(a) and \ref{fig:exactcase}(b). This directly yields a high-accuracy reference solution for $\Sigma = G/4$.

To simulate the output of a Monte Carlo-based impurity solver in imaginary time, we evaluate $G(\tau)$ at a large collection of random points $\tau$, and perturb the result by random noise of magnitude $\eta$. We then obtain a DLR expansion of $G$ by least squares fitting this data. This expansion can be evaluated in the Matsubara frequency domain using \eqref{eq:dlrexpansionmf}. In this case, we have $\omax = 1$, so we use the DLR cutoff parameter $\Lambda = \beta$, and set the DLR accuracy parameter to $\epsilon = \eta$. We calculate the leading coefficients of the asymptotic expansion of $\Sigma$ analytically: $\Sigma_0 = 0$, and $\Sigma_1 = 1/4$.
We then obtain $\Sigma$ using both the direct formula \eqref{eqn:dyson} and the CRM procedure \eqref{eq:minimdisc}, as shown in Fig. \ref{fig:exactcase}(c) for $\beta D = 100$ and $\eta = 10^{-4}$. The pointwise error of $\Sigma(i \nu_n)$, computed using the known result, is shown for both methods at $\eta = 10^{-4}$, $10^{-6}$, and $10^{-8}$ in Fig. \ref{fig:exactcase}(d). We observe the $\mathcal{O}(\eta \, \nu_{n}^{2})$ leading order error growth of the self-energy obtained from \eqref{eqn:dyson}. The CRM approach yields uniform accuracy at the noise level $\eta$ until the asymptotic region, at which point the correctly-captured leading-order asymptotics yields an error decreasing at the expected $\mathcal{O}(\nu_n^{-2})$ rate.

We note that even if $G$ were obtained with very little noise, the high frequency noise amplification eventually leads to incorrect tail behavior using \eqref{eqn:dyson}, whereas CRM yields a DLR expansion of $\Sigma$ with the correct asymptotic behavior on the full Matsubara frequency domain.

\section{\label{sec:results}Numerical examples}
We demonstrate the CRM method for two prototypical DMFT calculations: the interacting Bethe lattice and a three-band Hubbard model of Sr$_2$RuO$_4$.

\begin{figure*}
    \centering
    \includegraphics[width=2\columnwidth]{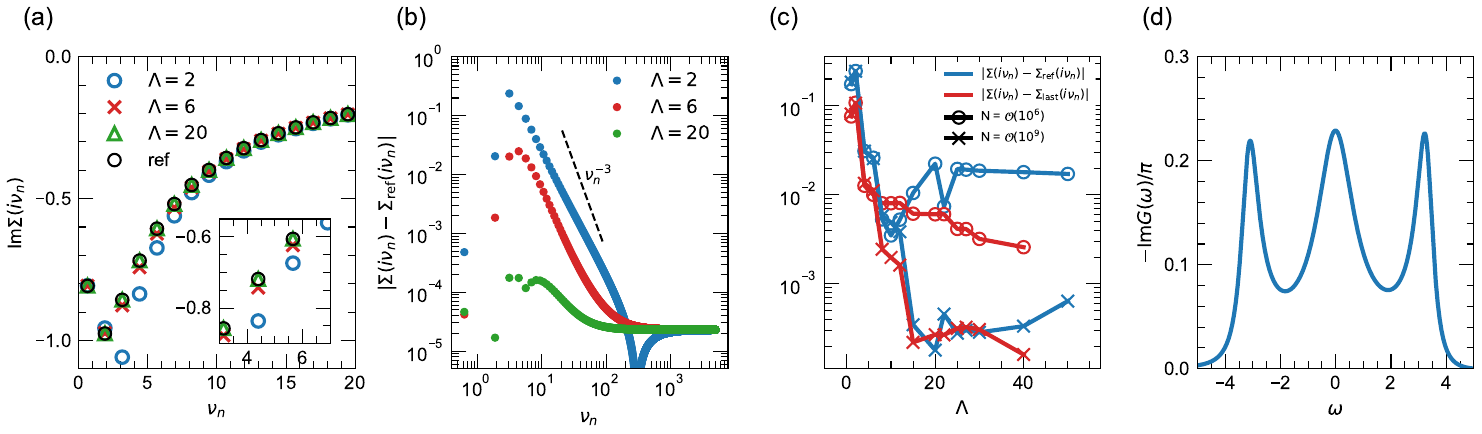}
    \caption{Interacting Bethe lattice for $U/D = 4$ and $\beta D = 5$. (a) $\Im \Sigma(i \nu_n)$ obtained usinl CRM for several choices of the DLR cutoff parameter $\Lambda$, with DLR tolerance $\epsilon = 10^{-6}$ and $N = 10^9$ QMC samples, along with the reference $\Sigma_{\text{ref}}$ (see text for details). (b) Pointwise absolute difference of $\Sigma$ and $\Sigma_{\text{ref}}$ from (a). (c) Maximum absolute difference of $\Sigma$ and $\Sigma_{\text{ref}}$, along with self-convergence error of $\Sigma$ versus $\Lambda$, for $N = 10^6$ and $10^9$ QMC samples. (d) Spectral function $-\Im G(\omega)/\pi$ approximated using Pad\'e analytic continuation of the Green's function.}
 \label{fig:bethe}
 \end{figure*}

\subsection{Interacting Bethe lattice} 
We begin with the half-filled, single-orbital Bethe lattice, for which the non-interacting Green's function is determined by a semi-circular density of states (DOS). The interaction Hamiltonian is of the simple form $\mathcal{H} = Un_{\uparrow}n_{\downarrow}$, where we have set (in units of the half-bandwidth $D=1$ eV) $U/D = 4$, and the system inverse temperature to $\beta D = 5$. We solve the corresponding quantum impurity problem using the CT-HYB impurity solver as implemented in TRIQS/cthyb \cite{Priyanka2016cthyb}. In order to simplify the experiment, we perform only one DMFT cycle, which is sufficient to test the CRM method for various choices of calculation parameters. The density matrix $\rho_{\text{imp}}$ is measured within CT-HYB, which allows us to compute the expectation values for the high-frequency expansion coefficients of the self-energy given in \eqref{eq:moments}.

In TRIQS/cthyb, samples of the imaginary time Green's function are averaged within bins of a fixed width, leading to a bin width parameter and a corresponding binning error. In our experiments, we use $10^{4}$ equispaced bins, which is sufficient to make the binning error negligible and to ensure that each bin contains a sufficient number of samples. We note that properly adjusting the number of bins requires some care, and an important future problem is to design an algorithm which maintains the efficiency of binning with substantially reduced errors, e.g., via projection onto a higher-order basis.
For all experiments, we set the DLR tolerance to $\epsilon = 10^{-6}$, which is below the Monte Carlo error, and we use CRM to obtain the impurity self-energy. We compare our results to a reference self-energy obtained from a numerical renormalization group (NRG) calculation. The NRG data was obtained using an NRG toolbox \cite{nrg1,nrg2,nrg3} based on the QSpace tensor library \cite{nrg4}, employing the full density-matrix NRG \cite{nrg5,nrg6}, a symmetric improved estimator for the self-energy \cite{Kugler2022}, and z-averaging \cite{nrg8} with extrapolation in the number of z-shifts \cite{nrg9}. 
\begin{figure}
     \centering
     \includegraphics[width=\columnwidth]{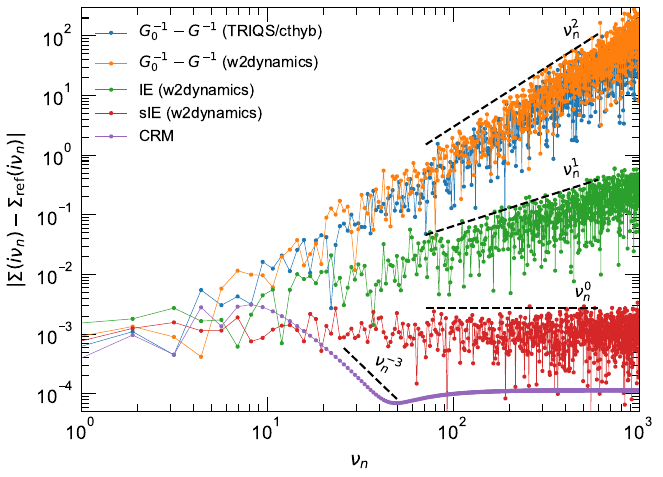}
     \caption{Comparison between constrained residual minimization (CRM), improved estimators (IE), and symmetric improved estimators (sIE) for the interacting Bethe lattice example. The CT-HYB solvers were run using $10^{8}$ QMC samples. For CRM, we have used $\Lambda = 20$ and $\varepsilon = 10^{-6}$. Dashed lines indicate power-law error behavior $\nu_{n}^{\alpha}$ for the different solutions: $\alpha = 2$ for the Dyson equation ($G_{0}^{-1}-G^{-1}$), $\alpha =1$ for IE, $\alpha = 0$ for sIE, and $\alpha= -3$ for CRM.} 
     \label{fig:cmp_int_bethe}
 \end{figure}

Using a CRM calculation with $N = 10^9$ QMC samples, in Fig. \ref{fig:bethe}(a) we plot $\Im \Sigma(i \nu_n) $ for several choices of the DLR cutoff parameter $\Lambda$, along with the reference. We plot their corresponding absolute pointwise difference in Fig. \ref{fig:bethe}(b). We observe convergence to the reference as $\Lambda$ is increased. 
\begin{figure*}
    \centering
    \includegraphics[width=2\columnwidth]{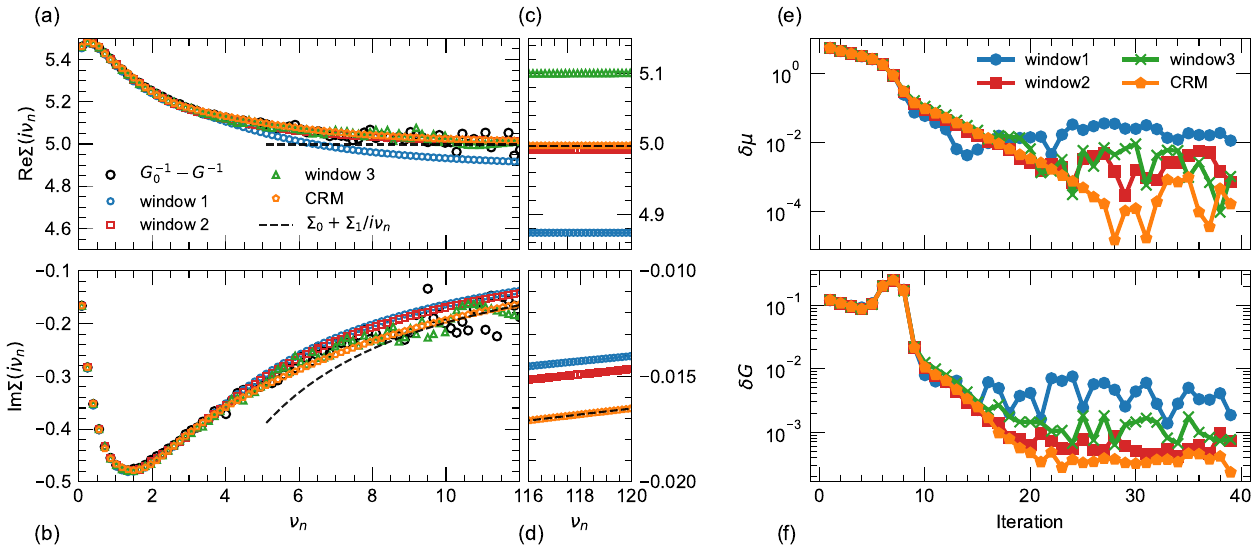}
    \caption{Comparison between the TRIQS tail-fitting algorithm and CRM for the Sr$_{2}$RuO$_{4}$ example. (a,b) Real and imaginary parts of $\Sigma^{11}(i\nu_{n})$ in the low to mid-frequency regime. The window sizes correspond to $(\nu_{n}^{\mathrm{min}}, \nu_{n}^{\mathrm{max}}) = $ (2,4) (window 1), (2,6) (window 2), and (2, 10) (window 3). The dashed black line corresponds to the directly measured high-frequency expansion of the self-energy. (c,d) Same as (a,b), at higher frequency. (e,f) Convergence of the chemical potential $\mu$ and local Green's function $G$ as measured by their differences $\delta \mu$ and $\delta G$ between subsequent iterations.} 
    \label{fig:sro}
\end{figure*}
Fig. \ref{fig:bethe}(c) shows the maximum absolute pointwise difference against $\Lambda$ (blue curves) for $N = 10^6$ and $N = 10^9$ QMC samples. We also show a ``self-convergence'' error (red curves), given by the maximum of $\abs{\Sigma(i \nu_n) - \Sigma_{\text{last}}(i \nu_n)}$, where $\Sigma$ is the self-energy computed with a given choice of $\Lambda$ and $\Sigma_{\text{last}}$ is the self-energy computed with the largest value of $\Lambda$ used, which is $\Lambda = 50$. Since in practice one does not have a known reference solution, the choice of a converged $\Lambda$ parameter can be determined by starting with a physically-motivated guess (based on an estimate of $\omax$), and plotting the self-convergence error against $\Lambda$ in this manner. Both error metrics consistently suggest convergence to the QMC noise level by $\Lambda = 20$, and this noise level decreases at the expected rate as the number of QMC samples is increased (by roughly a factor of $30$ for a $1000$-fold increase in the number of samples).
Fig. \ref{fig:bethe}(d) shows the spectral function (interacting DOS) obtained via Pad\'e analytic continuation \cite{FerrisPrabhu1973} for our single-orbital model. We find a spectral width of roughly $\omega_{\mathrm{max}} = 4 $ eV, consistent with the observed convergence near $\Lambda = \beta \omax = 20$.

We finally compare the CRM algorithm to the methods of improved estimators (IE) and symmetric improved estimators (sIE), in which the self-energy is obtained by measuring high-order correlation functions \cite{Hafermann2012improved, Hafermann2014, Gunacker2016worm, Kaufmann2019,Kugler2022}. 
We use the implementations of IE and sIE from the w2dynamics code~\cite{w2dynamics}. On a technical level, the IE/sIE requires an alternative Monte Carlo sampling method to obtain $\Sigma(i\nu_{n})$, compared to the standard sampling of $G(\tau)$ used for the Dyson equation~\cite{Gunacker2016worm, Kaufmann2019}. Therefore, to ensure uniformity across all calculations (and codes), the number of Monte Carlo samples in the different algorithms is adjusted to maintain (nearly) constant pointwise absolute error (relative to the NRG reference data) at the lowest Matsubara frequencies. 
Fig. \ref{fig:cmp_int_bethe} shows the pointwise absolute error of $\Sigma(i \nu_n)$ (compared with the reference $\Sigma_{\text{ref}}$) for each method, including the direct formula \eqref{eqn:dyson}. All methods agree in the low-frequency regime, and use of the direct formula produces the expected loss of accuracy at high frequencies. 

All three alternative methods outperform the direct formula \eqref{eqn:dyson}, but the $\mathcal{O}(\nu_n)$ error of IE at high frequencies is still too large for practical calculations. sIE and CRM outperform the other approaches, with constant error at high-frequencies, but CRM outperforms sIE in the high-frequency limit. This is a consequence of the higher accuracy obtained for the moments of the self-energy due to a lower Monte Carlo noise in the density matrix. Since our model is particle-hole symmetric, the high-frequency asymptotic expansion of $\Sigma(i \nu_n)$ does not contain an $\mathcal{O}(\nu_n^{-2})$ term; thus, the initial $\mathcal{O}(\nu_n^{-3})$ decay of the error indicates that the error of the constant part of the asymptotic expansion dominates the $\mathcal{O}(\nu_n^{-1})$ part. We note also that the sIE results come at a significant additional computational cost: for the simple one-band model considered here, the cost of measuring the sIE was roughly $10$ times larger 
than that of measuring of $G(\tau)$. By contrast, using CRM involves a negligible additional computational cost.

\subsection{Three-band Hubbard model for \texorpdfstring{Sr$_{2}$RuO$_{4}$}{Sr2RuO4}}
We next consider an {\it ab-initio}-derived three-band Hubbard model, a common DFT+DMFT workflow, at full DMFT self-consistency. We model the correlated Hund's metal Sr$_{2}$RuO$_{4}$, where the non-interacting low-energy physics consists of three partially filled $t_{2g} \equiv \{ d_{xy}, d_{xz}, d_{yz}\}$ bands from the Ru($4d$) site. We downfold the Kohn-Sham Hamiltonian onto three maximally localized Wannier functions, which serve as our correlated impurity subspace. Our low-energy model is identical to that discussed in Ref.~\onlinecite{Sheng2023}, which contains further details.

Interactions are governed by a local rotationally-invariant Hubbard-Kanamori Hamiltonian~\cite{Fujimori1995} with $U = 2.3$ eV and $J_{\mathrm{H}} = 0.4$ eV. We solve the impurity problem using the CT-HYB algorithm implemented in TRIQS/cthyb \cite{Priyanka2016cthyb}. We use the inverse system temperature $\beta = 40$ eV$^{-1}$ ($\sim$290 K). In each step of the self-consistent DMFT calculation, the CRM method is used to obtain the self-energy from the impurity solver. We also perform the same calculation using the standard TRIQS tail-fitting protocol, which operates according to the description in the introduction: an asymptotic expansion of a few terms is obtained by fitting the data produced by \eqref{eqn:dyson} on a user-specified frequency window, and glued to the result produced by \eqref{eqn:dyson}. We produce results for three different choices of the fitting window. For the CRM scheme, we take $\Lambda = 200$ and $\epsilon = 10^{-6}$.

The real and imaginary parts of the first diagonal element of the self-energy are shown in Figs.~\ref{fig:sro}(a)-(d), respectively, using CRM and the tail-fitting algorithm implemented in TRIQS, along with the direct formula \eqref{eqn:dyson}. The results for the other diagonal elements are similar, and the off-diagonal elements are zero. The self-energy produced by the CRM algorithm correctly reproduces the low-frequency result given by \eqref{eqn:dyson}, and has the correct high-frequency asymptotics, $\Re \Sigma(i\nu_{n}) \rightarrow \Sigma_{0}$ and $\Im \Sigma(i\nu_{n}) \sim \Sigma_{1}/(i\nu_{n})$, as imposed by CRM. 

The solutions produced using the tail-fitting method are highly dependent on the choice of the fitting window, which must be specified by the user. For too large of a window (window 3: $\nu_n^{\min} = 2$, $\nu_n^{\max} = 10$), too much high-frequency noise is incorporated, leading to large high-frequency errors. For too small of a window (window 1: $\nu_n^{\min} = 2$, $\nu_n^{\max} = 4$), the high-frequency asymptotics are not correctly captured. In a ``Goldilocks zone'' fitting window, it is possible to balance these competing sources of inaccuracy reasonably well, but in this experiment even a well-chosen window (window 2: $\nu_n^{\min} = 2$, $\nu_n^{\max} = 6$) is outperformed by CRM in reproducing the correct high-frequency asymptotics. Evidently, the necessity of finely hand-tuning the tail-fitting parameters---which must ideally be done for each non-equivalent orbital separately and, in principle, at each DMFT iteration--- represents a significant barrier to producing black-box DMFT codes, and CRM eliminates the need for such a procedure.

We also observe that the use of CRM helps to stabilize the DMFT loop by eliminating fluctuations in the high-frequency behavior of the self-energy.
In Figs. \ref{fig:sro}(e,f), we show the convergence during the DMFT iteration of two important quantities, the chemical potential $\mu$ and the local Green's function $G$, by plotting their differences $\delta \mu$ and $\delta G$ between consecutive iterations. The CRM approach appears to stabilize the DMFT self-consistency, leading to smoother and more rapid  convergence to a higher-accuracy result.

\section{\label{sec:conclusion}Summary}
We have proposed the CRM algorithm to compute the electronic self-energy in DMFT calculations using CT-QMC-based quantum impurity solvers in a stable manner. CRM is agnostic to the quantum impurity solver and can be used whenever the high-frequency behavior of the electronic self-energy is known, which includes applications outside of DMFT. We notelthat it might be possible to modify our method by relaxing the hard constraint in~\eqref{eq:minimdisc} to an adjustable penalty term set to roughly match the Monte Carlo noise level; however, we leave a study of the performance of this approach to our future research. The algorithm is simple to implement, and can therefore replace otherwise cumbersome tail-fitting procedures. Furthermore, our results suggest that CRM
yields improved stability and convergence of physical observables, and thereby improved accuracy of calculations within DMFT and DFT+DMFT. By combining the CRM method with the approach of Ref.~\onlinecite{Sheng2023}, which demonstrated the replacement of dense Matsubara frequency grids with compact grids of DLR nodes, we envision a streamlined and automated implementation of the DMFT loop with highly efficient discretizations of all quantities and minimal parameter-tuning. 

\section*{acknowledgements}
We are grateful to H. Strand for helpful discussions on alternative algorithms. We thank F. Kugler for useful discussions and providing relevant NRG data. We thank O. Parcollet, A. J. Millis, and N. Wentzell for their input. HL acknowledges support from grant NSF DMR-2045826. The Flatiron Institute is a division of the Simons Foundation.

\appendix

\section{\label{sec:app_high_freq} High-frequency asymptotics of the self-energy}

The high-frequency expansion of the interacting Green's function is given by
\begin{equation} \label{eq:gasymp}
   G(i\nu_{n}) = \frac{G_{1}}{i\nu_{n}} +  \frac{G_{2}}{(i\nu_{n})^{2}}+ \frac{G_{3}}{(i\nu_{n})^{3}} + \mathcal{O}\big(\nu_{n}^{-4}\big),
\end{equation}
with
\begin{equation*}
G_n = \partial_{\tau}^{n-1}G(0^{+})-\partial_{\tau}^{n-1}G(0^{-}).
\end{equation*}
The first few coefficients are given by \cite{Gull2011continuous}
\begin{equation} \label{eq:gcoefs}
    \begin{aligned}
        G_{1}^{ab} &= \langle \{d_{a},d_{b}^{\dagger}\}\rangle=\delta_{ab},\\
        G_{2}^{ab} &= -\langle \{ [\mathcal{H}, d_{a}], d_{b}^{\dagger}\}\rangle,\\
        G_{3}^{ab} &=\langle \{ [\mathcal{H}, d_{a}], [d_{b}^{\dagger}, \mathcal{H}]\}\rangle,
    \end{aligned}
\end{equation}
where $\mathcal{H}$ is defined in \eqref{eq:H}. The analogous expansion for the non-interacting Green's function is obtained using the Hamiltonian \eqref{eq:H} without the interaction term.

To obtain the high-frequency expansion of the self-energy, we substitute the expansions for the full and non-interacting Green's function into \eqref{eq:dyson1}. Matching terms of like powers yields
\begin{align*}
    \Sigma(i\nu_{n}) &= G_{2} - G_{02} + \frac{G_{3}-G_{03} + G_{02}^2 -G_2^2}{i \nu_n} + \mathcal{O}\big(\nu_{n}^{-2}\big) \\
    &\equiv \Sigma_0 + \frac{\Sigma_1}{i \nu_n} + \mathcal{O}\big(\nu_{n}^{-2}\big),
\end{align*}
where the coefficients of the asymptotic expansion of $G_0$ are given, in analogy with \eqref{eq:gasymp}, by $G_{01}$, $G_{02}$, etc.
The expressions for $\Sigma_{0}$ and $\Sigma_{1}$ in \eqref{eq:moments} then follow from \eqref{eq:gcoefs}.

\footnotetext[1]{The dimer example is part of the TRIQS benchmarks repository, which can be found at \url{https://github.com/TRIQS/benchmarks/blob/master/Dimer/notebook.ipynb}. The Jupyter notebook provides a description of the model and its implementation using the TRIQS software.}

\begin{figure}
    \centering
    \includegraphics[width=\columnwidth]{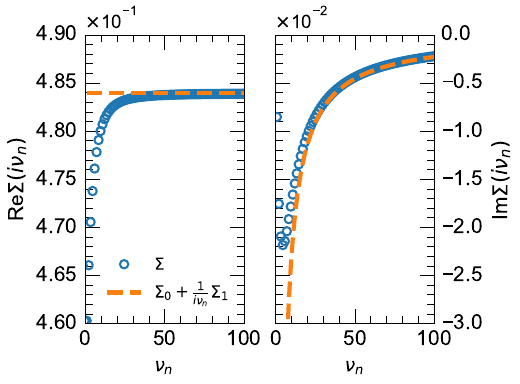}
    \caption{Asymptotic expansion $\Sigma_0 + \Sigma_1/(i \nu_n)$ of the self-energy for the dimer example, with coefficients computed from \eqref{eq:moments} using the measured many-body density matrix within CT-HYB, compared with a reference computed using exact diagonalization. 
    }
    \label{fig:dimermoments}
\end{figure}

We test our implementation on a simple model: a dimer coupled to a bath, which can be solved by exact diagonalization. A description of the model, calculation parameters, and implementation in TRIQS are provided in Ref.~\onlinecite{Note1}. 
We obtain a reference self-energy using the exact diagonalization library PyED \cite{Note2}. We use the TRIQS/cthyb \cite{Priyanka2016cthyb} solver to sample the many-body density matrix $\rho_{\text{imp}}$, from which we compute the first two coefficients of the asymptotic expansion of the self-energy using \eqref{eq:moments}. Fig. \ref{fig:dimermoments} compares an expansion $\Sigma_0 + \Sigma_1/(i \nu_n)$ to the reference self-energy, showing close agreement in the high-frequency regime. This implementation is available in latest release of TRIQS/cthyb~\cite{Note3}.

\footnotetext[2]{PyED: Exact diagonalization for finite quantum systems. \url{https://github.com/HugoStrand/pyed}.}
\footnotetext[3]{TRIQS/cthyb (v3.2.0) \url{https://github.com/TRIQS/cthyb/tree/3.2.x}}

\section{\texorpdfstring{$L^2(\tau)$}{L2} norm of a DLR expansion} \label{app:l2norm}
We show in this Appendix how to compute $||G||_{L^{2}(\tau)}$ for a DLR expansion $G$. In the CRM method, the objective function is given by $||R||_{L^{2}(\tau)}$, with $R$ represented by its values $r_k$ at the DLR Matsubara frequency nodes; the DLR coefficients $\wh{r}_l$ can be recovered via interpolation in the DLR basis. We focus on the case of scalar-valued $G$, since the matrix-valued case follows directly.

Let
\begin{equation}
G(\tau) = \sum_{l=1}^r K(\tau, \omega_l) \wh{g}_l
\end{equation}
be a DLR expansion. We have
\begin{align*}
||G||_{L^{2}(\tau)}^{2} &\equiv \frac{1}{\beta} \int_{0}^{\beta} d \tau \, G^{2}(\tau) \\ &= \frac{1}{\beta} \sum_{k,l=1}^r \hat{g}_{k}\hat{g}_{l}\int_{0}^{\beta}d\tau \, K(\tau, \omega_{k})K(\tau, \omega_{l}) \\ &\equiv \sum_{k,l=1}^r \wh{g}_k M_{kl} \wh{g}_l,
\end{align*}
with the last equivalence defining the matrix $M$. Thus the squared norm is given by a quadratic form with matrix $M$, evaluated at the vector of DLR coefficients. $M$ is straightforwardly computed, and we write two equivalent formulas:
\begin{equation} \label{eq:l2matrix1}
   M_{kl} = K(0, \omega_{k})K(0,\omega_{l}) \frac{1-e^{-\beta (\omega_k + \omega_l)}}{\beta (\omega_k + \omega_l)}
\end{equation}
and
\begin{equation} \label{eq:l2matrix2}
   M_{kl} = \frac{K(0,\omega_{k})K(0,\omega_{l}) - K(\beta, \omega_{k})K(\beta,\omega_{l})}{\beta(\omega_{k}+\omega_{l})}.
\end{equation}
The $\omega_k = -\omega_l$ case is given, via the limit, as $K(0, \omega_{k})K(0,\omega_{l})$.
When $\omega_k \approx -\omega_l$, both formulas suffer from catastrophic cancellation, but this is straightforwardly remedied in \eqref{eq:l2matrix1}, e.g., using the \texttt{expm1} function \cite{expm1} for the stable calculation of $e^x-1$. On the other hand, when $\beta(\omega_k + \omega_l) \ll 0$, \eqref{eq:l2matrix1} is vulnerable to numerical overflow, whereas \eqref{eq:l2matrix2} is numerically stable. We therefore use \eqref{eq:l2matrix1}, evaluated using the \texttt{expm1} function, when $0 < \abs{\beta (\omega_k + \omega_l)} < 1$, \eqref{eq:l2matrix2} when $\abs{\beta (\omega_k + \omega_l)} \geq 1$, and the exact formula when $\omega_k = -\omega_l$.

\bibliography{ref.bib}

\end{document}